# Workshops on Extreme Scale Design Automation (ESDA) Challenges and Opportunities for 2025 and Beyond

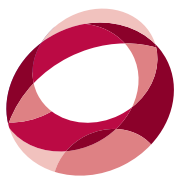

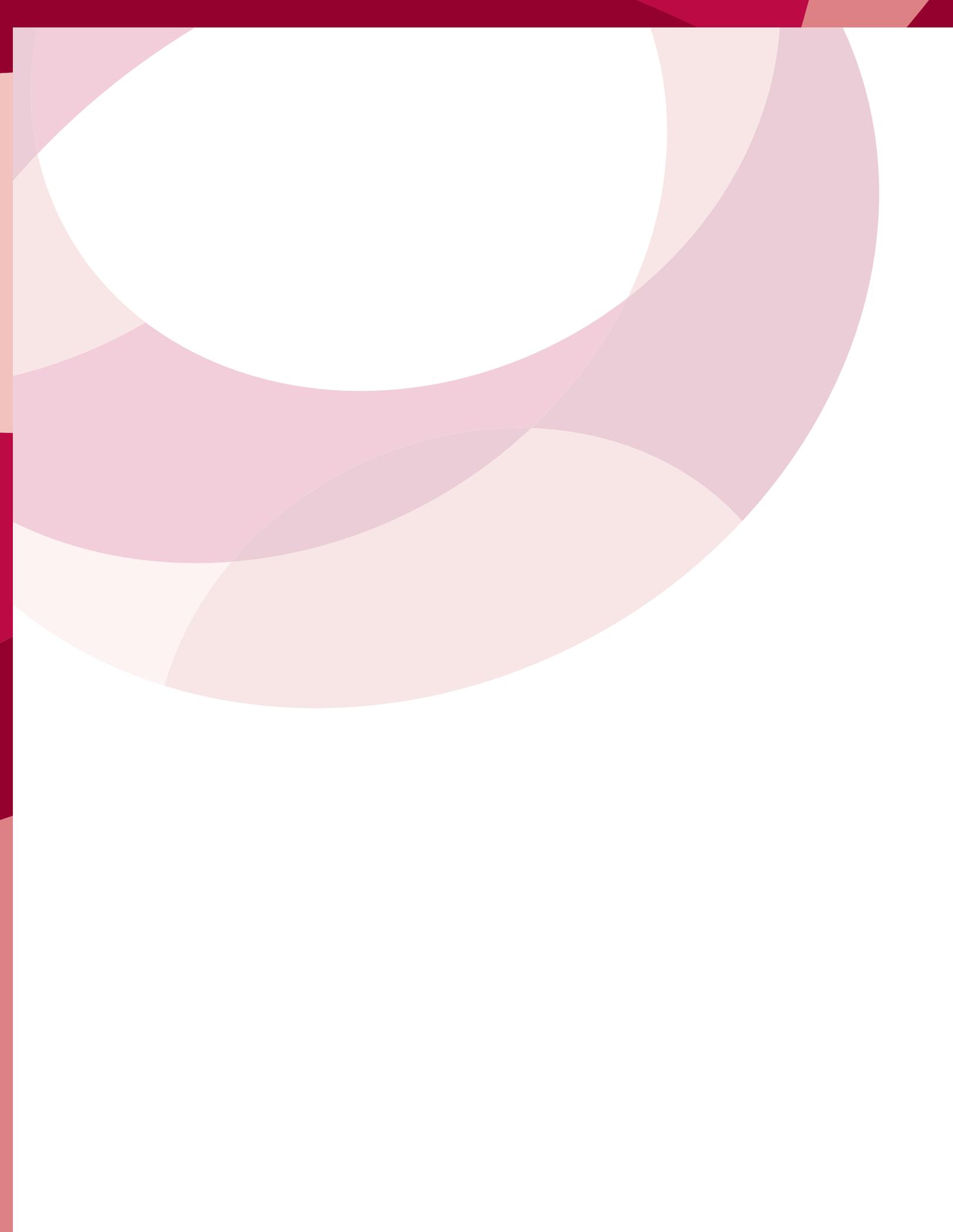

# Workshops on Extreme Scale Design Automation (ESDA) Challenges and Opportunities for 2025 and Beyond

R. Iris Bahar, Alex K. Jones, Srinivas Katkoori, Patrick H. Madden, Diana Marculescu, and Igor L. Markov

## Abstract

Integrated circuits and electronic systems, as well as design technologies, are evolving at a great rate—both quantitatively and qualitatively. Major developments include new interconnects and switching devices with atomic-scale uncertainty, the depth and scale of on-chip integration, electronic system-level integration, the increasing significance of software, as well as more effective means of design entry, compilation, algorithmic optimization, numerical simulation, pre- and post-silicon design validation, and chip test. Application targets and key markets are also shifting substantially from desktop CPUs to mobile platforms to an *Internet-of-Things* infrastructure. In light of these changes in electronic design contexts and given EDA's significant dependence on such context, the EDA community must adapt to these changes and focus on the opportunities for research and commercial success. The CCC workshop series on Extreme-Scale Design Automation, organized with the support of ACM SIGDA, studied challenges faced by the EDA community as well as new and exciting opportunities currently available. This document represents a summary of the findings from these meetings.

## Workshop Participants

R. Iris Bahar, Sankar Basu, Sanjukta Bhanja, Randy Bryant, Paul Bunyk, Krish Chakrabarty, Yiran Chen, Derek Chiou, Bob Colwell, Andre DeHon, Sujit Dey, Alex Doboli, Nik Dutt, Dale Edwards, Jim Faeder, Richard Goering, Patrick Groeneveld, Ian Harris, Mark Johnson, Alex Jones, Bill Joyner, Ramesh Karri, Srinivas Katkoori, Selcuk Kose, Steve Levitan, Hai Li, Xin Li, Patrick Madden, Diana Marculescu, Radu Marculescu, Igor L. Markov, Pinaki Mazumder, Mac McNamara, Noel Menezes, Prabhat Mishra, Natasa Miskov-Zivanov, Kartik Mohanram, Vijaykrishnan Narayanan, Sani Nassif, John Nestor, David Pan, Mandy Pant, Sudeep Pasricha, Rob Rutenbar, Sachin Sapatnekar, Lou Scheffer, Carl Sechen, Don Thomas, Josep Torrellas, Jacob White, Mehmet C. Yildiz, Hao Zheng.

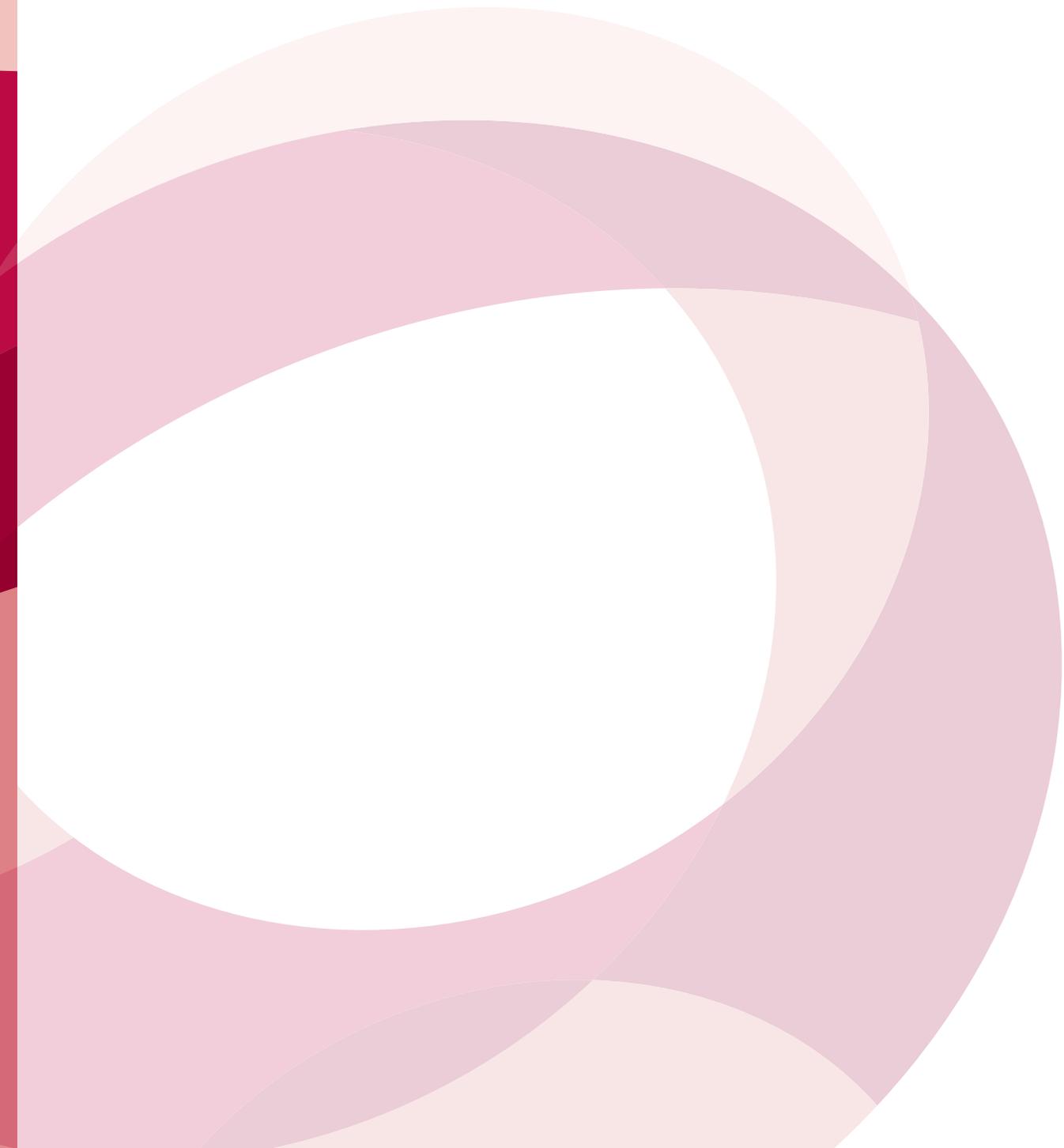

# Executive Summary

Integrated Circuit (IC) Technology and Electronic Design Automation (EDA) have reached a crossroads at the conclusion of Dennard scaling. The Extreme Scale Workshop Series was held to identify the best mechanisms to address the technology challenges and discuss the appropriate roadmap for Design Automation as a community. Consensus was reached on key topics and research directions categorized in three focus areas:

**Extreme-scale Electronic Design Automation** research is a *big-data* discipline that remains relevant and important to the US and World economies. Semiconductors, or electronics as a whole, are key economic and technology drivers. Even with slowed growth, the field indirectly supports several very large product categories and markets. EDA research must continue to enable the industry by addressing key challenges, such as system-level design, verification, and design for extreme scale devices. Of rising importance is closing the tool productivity gap for scaled technologies that currently dominate IC design starts.

**Emerging technologies** are being developed, but few are product-ready, and it is not clear yet which will be commercialized. The development of full system-design flows for these technologies should be a primary research focus. As in the early days of silicon EDA, effective abstractions can result from collaboration between technologists and EDA researchers to enable prototype tools, which can demonstrate successful new automation and lead to commercial products. Developing abstractions, benchmarks and metrics are essential first steps to making informed decisions about technologies to pursue for R&D investment.

**New markets** where Design Automation techniques and concepts can be applied will enable transformative advancements. Robust algorithms and heuristics can provide solutions to computationally complex, multi-objective problems based on technology abstractions. Areas where immediate investment is critical are cyber-physical and cyber-secure systems and include automotive, aerospace, robotic and energy applications. Longer-term directions include the development of wearable and implantable systems, as well as medical systems and technologies.

**Cross-cutting issues** include the development of appropriate abstractions, metrics and benchmarks in all of stated EDA focus areas, synergy of EDA with research in Computer Architecture, as well as the maintenance of a consistent EDA workforce. For example, as Computer Architectures trend to more specialized hardware, increasing embedded systems targets, and a growth in parallelism, Computer Architecture design teams will need to rely more heavily on automated tools. At the same time, the Design Automation field competes for new talent with several other exciting fields. Investment in EDA research can help invigorate the industry by attracting new professionals with unconventional thinking.

**The workshop participants identified the following specific actionable recommendations:**

◗ Lead the development of system-level design and verification techniques that enable highly competitive extreme-scale (e.g., $10^{15}$ devices) systems. Such techniques should include transformative high-level abstractions that abstract high-level behaviors, effectively model technology considerations, such as unreliable components, and capture relevant physical phenomena.

◗ Encourage improvements in algorithms, tools and methodologies for mature CMOS technology generations (i.e., >80nm). Many useful applications do not require advanced fabrication, but the design productivity gap and non-recurring costs remain challenges in such cases.

◗ In order to guide research in emerging technologies to be more predictable and conclusive, develop abstractions, benchmarks, and design metrics that would describe relevant challenges without unnecessary details. They will enable the development of design tools for automated design with these technologies, and for quantifying their potential impact.

◗ The development of design automation for electronics and its maturity has now enabled the *Design Automation of Things*. Design Automation techniques and methodologies should be translated to new areas when applicable. Further, Design Automation expertise should be combined with domain specific knowledge to enable new algorithms and methodologies for transformative technology advancements.

◗ Professional societies should take a more active leadership role in these efforts, especially toward transforming technology in markets beyond electronics. Holding regularly scheduled events to continuously refine a forward-looking vision for this community is an important thrust.

There are formidable challenges, tremendous opportunities, and great risks in how the end of Dennard scaling is handled. The EDA community must create an environment conducive of transformative results by identifying and addressing key design challenges of new markets. By moving cautiously yet decisively, it should maintain domestic leadership in current technologies as well as the technologies and markets of the future.

# 1  Introduction

The Electronic Design Automation (EDA) of very large-scale integrated (VLSI) circuits and systems has been a key enabler of the last half century's computing revolution. The abstraction of the physical properties of devices formed the basis of the EDA field which has in turn developed a rich set of tool flows to efficiently harness these devices. As the scaling of the underlying integrated circuit technology–often referred to as classical or Dennard scaling [1]–proceeded, the EDA community developed a new class of algorithms and optimization techniques. The powerful techniques that emerged continued to efficiently handle problems of increasing algorithmic complexity, integrated new optimization domains, and ultimately improved the capabilities of resulting integrated circuit designs. Thus, the last 40 years saw a golden age of scaling enabled by EDA that has dominated the advances in computing as predicted by Gordon Moore and colloquially referred to as *Moore's Law* [2]. This scaling is more tangibly exemplified by the progress since the Intel 4004 with just over 2,000 transistors in 1971 to the recent Intel Itanium line processors with more than two-billion transistors including Tukwilia (2010) and more recently Poulson (>3 billion transistors, 2012).

To leverage technology scaling, the capabilities of EDA tools have evolved dramatically over time. However, EDA tools still rely on the fundamental philosophy that has made the realization of billion transistor ICs possible. As one of the earliest inherently interdisciplinary fields, EDA has combined computing theory with applied mathematics, optimization, and the material sciences including physics and chemistry. The resulting abstractions of devices were used to create a separation of concerns that allow predictable and efficient processes for high-volume manufacturing of ICs. Starting with the lambda design rules of Mead and Conway [3], EDA has a history of allowing non-device technologists to effectively layout circuits. Over time the layers in the stack have increased, allowing designers to describe circuit behaviors at an algorithmic level, synthesize the resulting circuit, optimize the layout, and test for manufacturability while simulating and validating the result all at different levels of the flow. This powerful and scalable EDA methodology continues to provide IC designers and architects the means to harness Dennard scaling to construct powerful and sophisticated electronic systems.

Over time, this exponential growth (six orders of magnitude in approximately 40 years) has led to a trend of moving to higher levels of abstractions [4] to specify designs at the system level. Register-transfer level (RTL) specfication and synthesis have evolved to synthesis from software specfications written in high-level languages such as C++ and MATLAB. In current electronic design flows, the number of abstraction levels continues to grow, pushing higher into the system level and lower into artifacts of the further scaled technology (see Table 1). Unfortunately, these trends of increasingly numerous devices and capturing behaviors with higher levels of abstraction have led to a widening gap between what integrated circuits are physically capable of and how well the EDA tools can exploit these capabilities (see Figure 1). Even with the foresight and research investment by the EDA community to address this trend, the productivity gap remains a key challenge.

| Table 1: Modern Electronic Design: from applications to implementations ||
|---|---|
| 1. | Application domain, market, design platform |
| 2. | Product specifications |
| 3. | Software / programming |
| 4. | System architecture and microarchitecture |
| 5. | Functional and timing specifications |
| 6. | SoC view, chip floorplan |
| 7. | Design blocks and IP blocks |
| 8. | Constraints (timing, thermal, area, etc) |
| 9. | Design flow |
| 10. | Specific design tool(s) |
| 11. | Manufacturing technology, Process Design Kits |
| 12. | Dominant design concerns in a new technology |

Of the many technology challenges solved by EDA, the 80nm barrier was predicted at one point to be the physical fabrication limit. Today, projections are that beyond 5nm feature sizes do not look compelling, and even more near-term technologies require major changes such as FinFET transistors and relevant infrastructure. While many technology barriers have been overcome, each such leap aggravated the challenges for designers and EDA tools. The priority for the earliest EDA tools was to maximize what could be realized in a small die area. At some point, performance became the dominant metric. Sub-80nm technologies required careful control of power dissipation, especially due to current leakage. Technologies below 32nm add reliability concerns for transistors and interconnect. Such increasing design considerations (power, performance, cost, reliability), coupled with design complexity, have exacerbated the already widening productivity gap between tools and technology.

While EDA is faced with the challenges of new CMOS technology nodes, many new and emerging technologies are competing to augment and potentially replace silicon in an effort to continue Moore's Law. These devices and technologies require research investment into device models, abstractions, design tools, and validation mechanisms to enable their integration into hybrid CMOS flows. However, the EDA field itself has lost much of the excitement of the early years of continued innovation. The three largest EDA companies hold a dramatically high percentage of the $4 billion market while also having dramatically reduced their investment in research. Moreover, the naturally cyclic IC market experiences particularly severe peaks and valleys compared to other technology fields. Such market trends tend to disrupt the workforce pipeline severely. Ongoing hiring is focused on established and near-term expertise in areas such as place-and-route, low-power optimizations, hardware and software security, and cloud computing [5]. Furthermore, start-up companies no longer thrive in the EDA realm. Those few with useful technologies are often starved and eventually assimilated into one of the top-3

> **The purpose of this workshop series was to take an introspective look at the EDA field while crystalizing a vision for both the near and long term.**

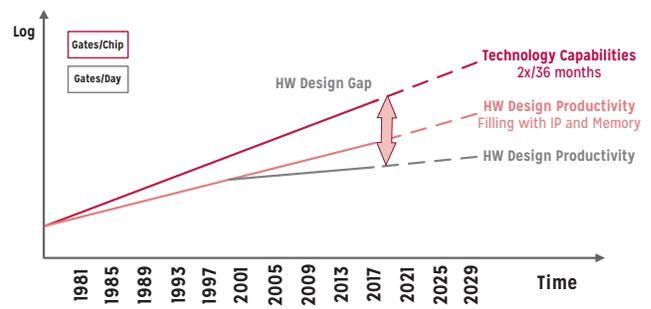

*Figure 1: The design productivity gap [6]*

companies, without providing rewards for new ideas and encouragement for further innovation.

As part of this process, the workshops examined both the successes and open challenges for EDA. Further, EDA needs were considered in the context of technology scaling and hybrid technology electronic systems. From these discussions, it became evident that the EDA field has and continues to develop a powerful and scalable toolkit of abstractions, algorithms, and design flows relevant to support and enable the design of current and future electronic ICs. However, this toolkit can be applied more broadly in the context of similarly complex problem domains in which abstractions are possible and algorithms for efficient design require heuristic solutions. In particular, the workshop series identified three key directions to achieve effective EDA development to the year 2025 and beyond:

**Extreme-Scale EDA:** EDA is perceived as a technically mature field where continued dedicated effort will only lead to modest progress in the field with limited impact. In contrast, the analysis suggests that critical and deep problems remain primarily unsolved. A focused effort to address the most relevant challenges has an opportunity for transformative impact. For example, verification of designs with billions of transistors is a grand challenge for EDA in the coming decade. Further, as scaling slows, there is an opportunity for EDA to explore methods to extract better results from existing technology nodes. The driver of Moore's law can be shifted from leveraging scaling to dramatic advancements from improved tools.

**EDA for Hybrid Post-CMOS Electronics:** Integration of a particular emerging technology may or may not bring about new ways of computing. For instance, carbon nanotubes and graphene attempt to emulate the functionality of CMOS, whereas memristors and quantum technologies have proposed different computing strategies such as implication logic [7] and cellular automata [8]. As a result, the integration of new and emerging technologies into EDA flows requires a re-examination of various levels of the traditional design stack from abstractions to verification.

**The Design Automation of Things**: The EDA field has developed an expertise in creating design abstractions for physical systems. These abstractions allow the development of algorithms to assist in the design of systems with billions of elements by developing highly effective and efficient heuristics that solve mildly non-linear, yet often NP-complete problems. These solutions can potentially benefit a variety of fields outside of the integrated circuit communities including cyber-physical systems, cyber-secure systems, and biology/medical technology.

Critical to the success of these directions is a renewed investment in the collaboration between researchers in industry and academia. Industry is well-tooled to address many near-term challenges in EDA flows. Unfortunately, the reduction of research investment in the EDA industry may result in important medium and long term challenges receiving less attention. Academics can provide this support but must work closely with the EDA industry to better understand the problems to be solved and to obtain relevant benchmarks and metrics. Further, a healthy investment in academic research is necessary to maintain an appropriate pipeline of professionals into the EDA industry. However, academics too must change the way they market and present EDA material to students for better recruiting into EDA. Together, industry and academia must reverse the growing perception of EDA as an unexciting, entirely mature, and comparatively difficult field in computer science and engineering. Figure 2 provides an overview of the main challenges for the EDA community and how we are proposing they should be addressed.

The remainder of this document focuses on three proposed thrust areas for investment to enable the successful design of electronics through EDA in the next ten years (through 2025) as well as strategies looking toward enabling the long term success of design practices of electronics and other relevant technologies. Section 2 reviews traditional EDA and highlights the deep and significant challenges that remain in this sub-field. Section 3 discusses emerging technologies for integrated electronics with computing and communication as drivers for EDA research and development. Section 4 focuses on new markets that can benefit from techniques developed for EDA. In particular, the EDA expertise (e.g., in dealing with practical large-scale systems) that can be applied to a broader range of applications are enumerated. Section 5 describes in more detail the cross-cutting challenges that affect each of these thrusts for EDA including 1) the need for appropriate abstractions, metrics and benchmarks, 2) the educational efforts in EDA that are necessary, and 3) the need for close interactions between the fields of EDA and computer architecture, which share some important goals and drivers. Finally, Section 6 presents conclusions and recommendations.

|  | **TRADITIONAL EDA** | **POST-CMOS** | **DESIGN AUTOMATION OF THINGS** |
|---|---|---|---|
| **What?** | Conventional CMOS device design will remain challenging, and the market is unlikely to shrink. | New technologies (CNT, spintronics, memristors, etc) may augment or even potentially replace CMOS in some areas. | Modern electronic devices are complex hybrid systems, integrating electronics and software. |
| **Why?** | Even with slowing adoption of new technology nodes, there are still many traditional EDA problems that are not solved adequately. | With decreasing gains from conventional CMOS, it is more critical than ever to explore new technologies. | A well designed system can be much more than the sum of its parts. The EDA community has a long history of dealing with large, complex design problems with competing objectives. |
| **How?** | To maintain the pipeline of new talent into traditional EDA areas, broader outreach, new ideas such as EDA MOOCs, and making EDA more visible are essential. | To ensure timely research, evaluation, adoption, and effective reuse of results, EDA abstractions, design metrics and benchmarks are needed. | Cross-disciplinary collaboration, between software developers, circuit designers, and technology developers, is essential. |

*Figure 2: An overview of the main challenges to be addressed by the EDA community (what), the reason these challenges are important thrusts of future EDA research (why), and initial steps required to address these challenges (how).*

## 2  Continuing Onward: Next-Generation Electronic Systems

In the early days of electronic design, tool developers were required to have an intimate understanding of the underlying technology to implement ICs, and EDA researchers had to be familiar with design methodologies of the day, as EDA software *tools* were intended for use by IC *designers*. For academic EDA researchers, these observations were crucial to ensure relevance of their research programs. EDA graduate students are still often encouraged or required to take architecture and VLSI design courses towards their degrees. But over time, as IC design methodology matured into a common flow with well-understood components (tools for synthesis, place-and-route, verification, etc.), it became possible for EDA researchers to successfully carry out their research independently on problem statements abstracted by the previous generation of design-centric EDA researchers. In some sense, the second and third generation of researchers stood on the shoulders of giants from the first EDA epoch [10]. This era of EDA research has been extremely successful in that we are able to design complex IC products today based on *point tools* commercialized as a result of decades of research by hundreds, if not thousands, of EDA researchers.

**The situation today is different:**

◗ Due to Moore's law scaling, IC products are complex systems leading to a vastly expanded set of system-level research problems.

◗ Embedded software, traditionally outside the mainstream of EDA research, plays a larger role in IC product development with the software team size often dominating the number of hardware engineers.

◗ IC product features are driven more by platform-level considerations due to power and thermal constraints. The role of displays, memories, power management IC's, thermal sensors, and packaging at the platform level need to be well-understood for IC design.

◗ The rapid pace of product introduction driven by end-user demand has led to drastically shorter design cycles thereby increasing the reliance on EDA.

◗ Process technology complexities need to be comprehended at the system level for the right tradeoff between power, performance, and reliability.

> **This uncertain, and exciting environment is reminiscent of the beginnings of the EDA era prior to the stability provided by Dennard scaling.**

This time, the uncertainty of the current environment is alleviated by the large global demand for the IC products that EDA enables along with the absolute necessity of research along new system-level directions. While interactions between EDA research and IC design are beneficial, the levels of complexity in each field also make these interactions challenging, thus hampering potential advancements. These complexity challenges also discourage computer science students, whose help is necessary to develop embedded and processor-specific software, from entering the EDA field (see Section 5.3). Aligning with systems engineering architects may be particularly beneficial for EDA developers, as this brings in a rich new set of interesting and important problems to pursue.

In the remainder of this section we review the challenges presented by these trends to traditional EDA flows. Further, we discuss the need for a renewed emphasis on appropriate abstractions, metrics, and benchmarks to promote a collaborative industry and academic effort to achieve success. Finally, we focus on new directions for "big data" approaches that are required to enable EDA of systems and ensure its scalability.

## 2.1 The Role and Promise of Traditional EDA Research

Traditional EDA can be defined as software supporting design flows that implement a single IC with traditional silicon fabrication, but not an entire system or software that runs on the system. Standard EDA steps include: (i) synthesis, placement, and routing, (ii) RTL verification, (iii) reliability verification, (iv) electrical simulation. Multi-objective optimization, verification, and test can now be considered traditional. But such a classification does not explicitly require details of the target technology node. As a result, modern electronic design has become a balancing act between several key design metrics, relevance to upcoming applications, the complexity of the design process, and time-to-market. Such tradeoffs are illustrated by the spider chart in Figure 3. While EDA's critical role in electronic design is widely acknowledged, the value of EDA, specific investment targets and the room for growth in EDA, continue to be a source for debate. There are two prevailing opinions: (1) a research focus is still required for the traditional EDA core disciplines, and (2) research should focus on higher-level concerns of system-level design and embedded software. Hence, we discuss challenges relevant to success in these directions.

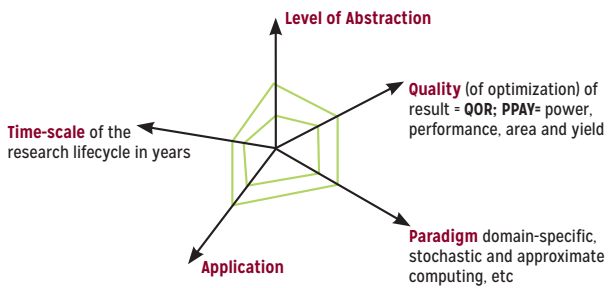

Figure 3: Key aspects of modern research and development in EDA.

**Core EDA Research and the Impact of Further Improvements:** Published literature has and continues to demonstrate substantial advances in EDA for core tool capabilities. However, more importantly, recent, essentially academic, contests have demonstrated powerful and high-quality tools and algorithms that are not available through EDA vendors. Best-performing teams at recent contests have pushed the limits in terms of runtime and solution quality. IBM, Intel, and Mentor Graphics are among those EDA industry users and vendors that have benefited from results of contests in finding new ideas for their own internal tools. However, vendors have been either unable or unwilling to adopt these results into their own tool chains, even when algorithms are clearly described in publications and not difficult to reimplement. It seems that EDA vendors are not convinced by the risk-reward tradeoff of such involvement. The customers of these tools already have a "good enough" solution, whereas possible improvements and impact are not widely understood and sometimes dismissed. Moreover, while *"push-button"* flows (promised by vendors ten years ago) may still be achievable with adequate investment, the current climate precludes this advancement.

Perhaps the greatest obstacles to adapting university research in industry have been the feature complexity of practical IC designs as well as the fear of unintended side-effects and disruptions. To illustrate this, somewhat surprisingly, IC designers sometimes forego readily-available optimizations in industry grade tools that could improve the quality of their results. An important reason behind this decision is the aggressive pursuit of a reduced time-to-market. Even very fast optimizations with high quality results increase the risk of introducing correctness problems or otherwise destabilizing the design flow. With relatively small design teams, these risks often outweigh the potential rewards of a particular optimization provided by the tools. However, soon these sub-optimal solutions will no longer be good enough, particularly with regard to thermal and power concerns. Thermal characteristics of ICs have improved dramatically, but battery life and power density remain important and are likely to benefit from further improvements in EDA tools. Similarly, while the most effective power reductions can be achieved through software optimizations by turning off unnecessary consumers of power, simultaneous optimizations across several layers of design abstractions have not been sufficiently explored, yet promise further necessary improvements.

**Scenarios for Leveraging Core EDA Research:** In some important cases, improvements spearheaded by academic tools may enjoy direct, dramatic, and broad impacts. Fewer designs target the newest technology nodes today due to high costs and limited access to these design technologies. Even for large-volume designs that can justify the most advanced manufacturing, yield and reliability concerns raise serious questions. Therefore, an increasing fraction of design starts target older technology nodes, with 130nm designs becoming increasingly popular. This can drastically reduce upfront costs and uncertainties in the design flow, while increasing yield.

> **An EDA advancement demonstrated in a tool that finds superior solutions is particularly valuable to effectively utilize legacy technology nodes.**

Thus, we can compare the added value provided by semiconductor scaling to quality improvements delivered by the EDA tools.

A different scenario in which core EDA research and expertise are often critically important is the support for research in VLSI design. Such support is required to demonstrate and properly evaluate new ideas in VLSI design (such as power reduction by using adaptive body bias for transistors, or diminishing the impact of process variation on clock trees by using tunable delay buffers). However, design researchers lack such an expertise while EDA researchers are not sufficiently rewarded to play a supporting role. Without a synergy, each part has to rely on its own resources and skills, thus limiting possible research accomplishments. Perhaps, federal funding agencies can identify an effective formula to encourage effective collaborations.

**A Data-centric Approach to System-level Design and Software:** At the 2009 NSF workshop on EDA [11, 12], a strong case was made by EDA industry to focus on *design data* rather than algorithms. This suggestion was reiterated in the CCC workshop series by additional industry representatives and academics, with the added emphasis on *structure in data*. Moreover, large companies are forced to focus on the system-level because silicon alone does not provide sufficient functionality, feature differentiation, flexibility, and future-proof properties. This expansion of concerns dramatically increases the amount of design data handled by EDA flows.

EDA researchers could tackle current challenges along two lines: (1) structure the design data by developing frameworks and infrastructures to enable the most effective operations, or (2) derive useful structured information using data-mining techniques. Effectively, EDA for 2025 and beyond has become a grand challenge in the *Big Data* domain. To this end, Gartner estimates that half of all data stored today were produced in the last two years, and 4.4M jobs will be in this big data domain by 2015 [13]. The next ten years must see solutions that coalesce traditional EDA techniques, many of which already deal with big datasets, with software practices designed to handle and manipulate big data [14]. For example, *machine learning* deserves exploration in this context. As we move into less-traditional EDA domains including platform-level optimization, post-silicon validation and embedded software, discovering the most effective dataset structure for specific workflow tasks becomes an important direction in EDA research. Despite massive amounts of data produced by existing flows, deriving actionable semantic information remains challenging. This is especially true for system validation.

**The Increasing Significance of Software:** Given the trends, current design projects often employ more software engineers than hardware engineers. This aspect of the electronics industry is illustrated by the Synopsys acquisition of Coverity, a company focused entirely on software, to supplement their design flow. More effective strategies are needed for integrating traditional EDA with software-based strategies in system or platform-level design flows. There are also significant implications to workforce training–software skills are often required in addition to traditional training in electrical or computer engineering.

## 2.2 The Role of Abstractions, Design Metrics, EDA Benchmarks, and Computational Experiments

The complexity of EDA has traditionally been addressed by establishing multiple levels of abstraction, developing extensive modeling and computational techniques relevant to each level, and developing a means of connecting different levels of abstraction. Additionally, powerful abstractions support cross-cutting research and development by facilitating reuse of results and techniques between different fields.

**Abstractions:** By an EDA *abstraction*, we mean a mathematically-specified self-contained model that omits a large amount of application detail, but expresses the most pertinent aspects of a design or design challenge. Common examples include the *logic circuit* abstraction and the *floorplan* abstraction. Abstraction-based EDA has been crucial to both academic research and industrial development. Many traditional EDA abstractions, such as *standard-cell layout*, remain largely relevant today, even when the focus of design optimizations changes. Other abstractions, such as *compact delay models*, closely follow technology and have a shorter half-life. Significant efforts have been invested into the validation of abstractions in practice, as well as their enhancement. Such efforts are necessary to maintain the relevance of abstractions used in ongoing work.

**EDA Benchmarks:** For EDA research to be relevant and applicable to industry tool-flows it must be evaluated with *representative benchmarks*. These benchmarks must also evolve with advances in manufacturing and design technologies, and reflect changes of focus for EDA optimizations. Recent benchmark releases by IBM, Intel, Mentor Graphics, and other companies illustrate how this can be accomplished [15, 16, 17].

**Design Metrics:** Along with abstractions and benchmarks, it is important to develop *design metrics* that correlate with practical quality-of-results (QOR) evaluation. In addition to connecting abstractions to applications, such metrics give graduate students and researchers concrete optimization objectives that can be used to compare competing approaches. This allows young researchers to learn and become productive more quickly, and reduces the need to fully understand the entire design flow to become productive.

**Computational Experiments:** serve to evaluate optimization methods with respect to design metrics using benchmarks. They must ensure the statistical significance of results, support the reproducibility by other researchers, and help determine how closely a given abstraction matches the reality of practical IC design.

VLSI researchers have been creating abstractions for a long time, including those for analog and mixed-signal circuits, large IC blocks, etc. However, existing abstractions are insufficient. In particular, analog design remains challenging to automate precisely because it is hard to create comprehensive and useful abstractions. System-level design and verification suffer similar difficulties.

In this context, we would like to encourage research that can develop new, effective abstractions and evaluate their adequacy in specific applications. Like in the early days of EDA research, abstractions can be enabling and economically effective due to *(i)* larger problem sizes, *(ii)* many new technology concerns and features, and *(iii)* the confluence of low-level and high-level aspects, as well as diverse technologies involved simultaneously. Desirable qualities of abstractions include: accuracy, scalability, succinctness, simplicity, clean semantics, changeability/composability, and amenability to analysis, computation and fast updates. With these criteria in mind, *it is critical to elevate abstractions to the status of important research results*, whereas so far abstractions have been considered intermediate results. Not to be purely theoretical, new abstractions should be made applicable and impactful through the use of design metrics, benchmarks, and computational experiments.

## 2.3 Rebuilding Interactions Between Academic Research and the EDA Industry

The level of technical interaction between academia and EDA industry has significant room for improvement. This is particularly surprising in 2013-14, as long-term research activities in EDA companies are at a very low level, and academic research can help to fill the void.

EDA companies focus on product development and are often conservative with respect to research investment. Rather than develop new ideas internally, an EDA company might prefer to purchase a start-up. There is also an inclination to broadly dismiss academic research for lacking access to key problems or industry-grade designs. Ironically, these same companies are often unaware of the most relevant academic research. Even when a research idea is known, it may still be dismissed as too preliminary, or it may look too difficult to implement in an industry-grade tool-flow. Some of these perceptions stem from stagnant EDA revenues, undermining funding for innovation in company labs and academic groups.

An apparent solution is to develop long-term strategies that increase collaboration between academic and industrial partners, particularly in the newest and least developed research directions. EDA companies should follow the most recent research results and identify those that have the most potential to benefit the company. Reaching out to researchers even with a small investment would provide encouragement and feedback on how to make their research more relevant to the industry. To provide an incentive for academics to follow this guidance, eventually there must be a reward such as paths towards publication and/or research funding. The joint-funding model of programs combining SRC's technical vision with NSF's scholarly review process is valuable to support this thrust. It is a modest goal to revitalize these collaborations and prompt the EDA industry to directly fund academic research. The NSF Industry/University Collaborative Research Centers (I/UCRC) also provide a valuable model of funding.

## 3  A Changing Landscape: New Technologies for Integrated Circuits

A number of promising technologies have emerged over the past few years. While some are still in their infancy, others have gained traction and some appear in commercial products. It is still unclear which of these technologies will emerge as success stories for computing hardware or if any of them can displace silicon-based technologies. Nevertheless, there will still be a need in the near- and mid- term for EDA tools that aid in the design, analysis, and optimization of such devices in future computing systems. In Figure 4 we list some of these technologies without indicating preferences.

### 3.1  Challenges for Designs with New Technologies

Regardless of which technologies actually take hold, some issues familiar from traditional EDA will remain important for emerging technologies. That is, abstractions, EDA benchmarks, design metrics, and computational experiments mentioned in Section 2.2 will play an important role in validating and eventually enabling widespread use of these new technologies. However, compared to silicon-based EDA, there will be greater emphasis on device models and appropriate simulation technology. In particular, a completely different method to modeling these devices may be necessary to capture appropriate behavior. This modeling effort will be critically important to accurately evaluate performance and enable fair comparisons against traditional designs. Additionally, several fundamental challenges affect multiple emerging technologies and will need to be addressed from the EDA perspective. Not necessarily unique to emerging technologies, these challenges are more pressing and may block future adoption in mainstream electronics.

**Integration:** A new technology may require a fundamentally different approach to coordinate devices and data to create the system. There will also be a need to allow for heterogeneous integration of modeling and simulation tools for diverse technologies into one system (or system-on-chip).

**Variability/Reliability:** Variability and reliability will be major issues with many of the emerging technologies. Reliability and security need to be part of the whole EDA synthesis and analysis process. The system may need to be evaluated from a probabilistic viewpoint.

**Big Data:** The amount of data that needs to be

**10**

| TECHNOLOGY | PROPOSED USE |
|---|---|
| Optical interconnect, optical devices | High-performance, high-bandwidth communications |
| Terahertz (RF) circuits | Automotive radar, security, high-bandwidth wireless communcation |
| Microelectromechanical systems (MEMS) and Nanoelectromechanical (NEMS) | Mechanical filter/switches, wideband antennas, gyroscopres, energy harvesting, data storage, sensors |
| Spintronics/multiferroics | Modeling synapse, physical brain/biomimetric behavior |
| Flexible electronics | Wearable computing, body tracking, glucose monitoring |
| Qbit technologies | Quantum computing/annealing/optimization |
| Phase-change memories (including memristors) | DNA memory (having long retention) |
| Microfludicis | Lab-on-a-chip, cooling |
| Steep slope devices | Ultra low-power computing |
| Superconductors | Ultra low-power computing, ultra high performance |
| Carbon-based electronics | Ultra low-power computing, high performance, monolithic 3D ICs |

*Figure 4: Emerging technologies for the EDA community.*

processed in order to evaluate and understand low-level (e.g., atomistic) behavior and to explore all reachable system states is huge. There may need to be better statistical methods for simulations involving big data.

**Flexible Models:** There will be a need for more flexible, modular, and/or extendable tools that can easily incorporate different behavior from various emerging technologies. A building-block approach, where each block is easily modifiable may be particularly advantageous.

**High-level Abstractions:** The benefits of high-level prototyping, estimation, synthesis, and verification can only be unlocked with appropriate high-level design and EDA abstractions. Some emerging technologies are particularly dependent on new abstractions and models– from logic to system level.

**Physical Layout:** New devices and circuit styles typically require rethinking traditional notions of physical layout, synthesis, extraction, and verification.

## 3.2 New Design Processes for New Technologies

Focusing on high-level analysis may require rethinking how systems are specified, designed, and implemented. If we take a current system that has been optimally designed using traditional silicon technology, and then simply replace each silicon component with an emerging-technology-equivalent, it may not yield any significant benefits (and may even be worse). This may not be surprising, since design decisions at the highest level are often the result of constraints dictated at the technology level. Instead, if we have the capacity to accomplish design exploration at the highest level first, unconstrained by technology, we may end up with a completely new way of computing that could allow us to better exploit the best properties of new technologies. Only then will it become apparent what kind of design automation tools would be most beneficial for these new technologies. We need new models that give designers the flexibility to change the high-level structure.

With these high-level analysis tools in place, we can also determine the feasibility of pursuing particular technologies on a large scale by considering *what-if* scenarios. If a technology is incompatible with high-level analysis, then there is no reason to develop low level detailed tools for more accurately analyzing physical effects. How these high-level analysis tools are constructed themselves will also be important. For instance, as a prototyping tool, a plug-and-play model could provide the modularity needed to allow for different levels of model abstractions to be swapped in and out of the tool. While EDA designers are very familiar with developing abstractions there is not necessarily a good exchange format available in order to go between levels.

While high-level analysis is appropriate at the early stages of development, physical implementation and technology-specific concerns must also be taken into account, which may require handling large amounts of data. The emphasis on big data for design automation has been recently amplified at the Design Automation Conference with a discussion of challenges and opportunities in managing big data for EDA [18]. Presenters from the industry identified significant progress and pointed out important bottlenecks in how distributed chip-design teams operate today. Attempts to leverage cloud computing and storage met with some success, but much of their promise remains to be exploited. Companies still express concerns about data security; however, restricting themselves to in-house computing severely limits the flexibility in how design efforts are organized, time-to-market, and overall team productivity.

Several stages of development are required for EDA to manage emerging technologies and realize their full potential, but researchers must be careful in identifying the most innovative and impactful research targets in each case. Given that the feasibility and the impact of new technologies cannot be reliably evaluated at first, they sometimes stir controversy in the research community. As a result, these technologies have a tendency to go through wide variations in expectations over time until their behaviors and potential uses are well-understood. As illustrated in Figure 5, initial acceptance often turns into exuberance, followed by skepticism, and eventually pragmatic adoption. While

### The Hype Cycle of Innovation

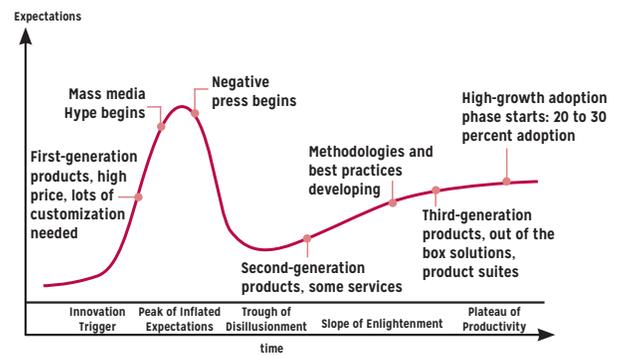

*Figure 5: A generalized pattern of research and commercialization– from conception to productive use [19].*

some of this hype and pessimism may be unavoidable, it is desirable to reduce the peaks and valleys and shorten the time to reach the "slope of enlightenment." If, for example, researchers view the physical demonstration of working novel devices and circuits as real science and the main result, while the development of tools that accurately model novel devices and facilitate large-scale circuits as an intermediate engineering task of smaller significance, this will not help in the eventual adoption of the new technology in the long run. Indeed, such cognitive biases and misconceptions can be damaging in terms of industry investment, funding priorities, final results, as well as recruiting and training eventual workforce participants. On the other hand, the emphasis on device models, design metrics, and benchmarks may facilitate pragmatic evaluation from the early stages, whereas the development of effective abstractions may increase eventual adoption by lowering barriers to entry and broadening reuse. To this end, funding agencies should encourage research proposals on new design technologies to develop an integrated EDA component and emphasize collaboration with EDA researchers. A clear, early understanding of key issues in scaling, synthesis algorithms, optimization, and verification would not only contribute to knowledge discovery, but also spearhead commercial adoption.

# 4 Looking Forward: New Markets and Applications

Design automation for silicon systems (traditional EDA) has been instrumental in achieving the exponential increases in device density and performance as predicted by Moore's law. Indeed, the unprecedented innovation afforded by the EDA tools and methodologies have made Moore's law (along with the resulting proliferation of electronics) a self-fulfilling prophecy. When thinking about the applicability of a similar approach to other markets, we must identify the unique features that have enabled EDA's successful quest.

## 4.1 Re-evaluating Design Automation as a Field

Traditional EDA has made feasible the process of analyzing complex problems and providing efficient solutions for them. Typically, these problems are hard to solve, either because of scale or because of the complexity of underlying phenomena and observable behavior. EDA has automated the design flow of integrated systems and made them *predictable* in simulation through the development of system-level models. Models, analysis, simulation, and verification have therefore been indispensable to streamlining the design process and meeting product specification(s) via efficient optimization.

When trying to assess the applicability of well-understood paradigms to newer markets, we must clearly identify potential benefits brought by design automation and their relevance to unsolved challenges in that domain. The strengths of silicon-centric EDA include the capability to separate and improve *design*, *construction*, and *optimization*. *Ad hoc* approaches in other markets can be made more rigorous through abstraction layers and associated predictive models, efficient algorithmic approaches, and resulting design flows—common characteristics for traditional EDA. Moreover, experts in many fields are focusing on tools, models, and abstractions. A distinct capability in EDA that can help in other markets is making sophisticated low-level behaviors amenable to high-level analysis, design, optimization, and verification through clear abstraction layers and predictive models. EDA is also unique in the scale of problem instances it solves-billions of separate objects (such as placeable cells and routable wires) can be handled on single-processor workstations. But what has made it truly different is its ability to open up design to non-experts with diverse backgrounds, creating large new markets. One can now start from high-level specifications and obtain a final design in a fully automated manner, something that was not possible without a design flow or deep knowledge of the circuit design process. In other fields, some of these challenges remain today: for example, the technologies for image capturing and analysis using iPhones and MRI machines are comparable, but iPhone implementations will end up being much cheaper. Can the cheaper technologies be made accessible to *non-experts* for the same kind of objectives (in this case, making high-quality medical diagnosis cheaper)?

We can conclude that design automation (DA) techniques can be beneficial in areas that:

◗ Need to be understood through analysis before being designed or constructed;

◗ Lack appropriate abstraction layers that can enable clean, predictive modeling;

◗ Must rely on both optimization and analysis to meet design specifications;

◗ Can make high cost technologies accessible via low cost alternatives for similar purpose;

◗ Need efficient assessment of expensive experimental outcomes beforehand.

We note that making this possible requires solving challenges not currently addressed by the existing DA body of knowledge. Of course, EDA has been effective for creating synergies between different areas and pulling together the models, analysis, and design of silicon systems. The complexity for new markets often comes from putting together components from vastly different fields, such as biology, medicine, material science, electrical engineering, and computer science. However, these interdisciplinary concerns are fundamentally similar to the early days of silicon. Applying the DA body of knowledge to new markets will require the development of models, abstractions, benchmarks, and associated simulation technologies that are able to assemble together and characterize components coming from different domains with vastly different dynamics and timescales. Of course, these abstractions and models cannot be identical to the early approaches of electronics design.

|   | MEDICAL TECHNOLOGY (*) | CYBER-PHYSICAL SYSTEMS (CPS) (*) | DESIGN FOR SECURITY (DFS) (*) | MISCELLANEOUS |
|---|---|---|---|---|
| I | • Synthetic biology, systems biology | • Automotive<br>• Datacenters (computing, cooling, interface w/ grid) | • Verification of hw/sw/protocols (and components thereof)<br>• Security and privacy in the cloud | • Hierarchical cloud design and applications |
| E | • Medical electronics<br>• Drug discovery | • Smart grid<br>• Renewable energy<br>• Design for (DF) wearable systems | • Trust verification | • Financial applications - "hedge fund engineering" |
| P | • Customized therapies<br>• Clinical diagnosis<br>• Prosthetics | • Robotics<br>• Industrial internet<br>• Internet of Things |   |   |

*Figure 6: Identified, Emerging, and Potential New Markets for DA*
*(\*) indicates a strategic priority*

For silicon, classic abstractions make design, optimization, and verification easier, but many other fields operate with abstractions and models that must include inherent uncertainty, arising from such factors as different information processing or storage substrates, sensing modalities, or interacting with the environment. Deep silicon scaling has required incorporating variability in modeling and analysis into the design process. These resulting techniques can be *exported* to other fields, but the type of uncertainty faced by systems in new markets may be very different and span multiple temporal or spatial scales. Incorporating uncertainty at varying scales, for a variety of observable metrics of interest, poses a challenge that is unlike those faced by traditional EDA early in its development. However, the DA expertise of addressing new design metrics can be useful in addressing such challenges.

## 4.2 Potential, Emerging, and Identified New Markets for DA

We have identified several new markets that can (or already have started to) benefit from the use of DA methodologies. Depending on where they are in terms of transferring DA knowledge to that field, new markets are marked as identified (**I**, i.e., sizable results already exist that show feasible application), emerging (**E**, i.e., evidence of applicability exists, but is still in development), and potential (**P**, i.e., no evidence exists, but possible application of DA would greatly enhance the field). Fields that should be considered strategic priorities are marked by (\*) in Figure 6. We also show in Figure 7 where the high priority directions are in terms of ease and readiness for transferring DA knowledge. Many of these research domains have already benefited from cross-fertilization with design automation. We list below a few of the recent funding efforts that have emphasized this cross-fertilization.

- A 2013 CCC visioning workshop on hardware security has featured a true design flavor and, consequently, was (unsurprisingly) sponsored by the SRC [20]. As a result of this visioning effort, NSF's CISE directorate has developed a crosscutting program on Secure and Trustworthy Cyberspace: Secure, Trustworthy, Assured, and Resilient Semiconductors and Systems (SaTC: STARSS) [21] which addresses some of the design, but few of the DA aspects mentioned herein.

- Furthermore, the 2008 NSF-sponsored workshop charting the path of the new (at the time) Cyber-Physical Systems (CPS) program [22] has identified DA tools and methods as being essential for CPS development–yet, they are conspicuously absent from existing calls for proposal for this program.

- Finally, in the context of bio-systems, the SRC has initiated a new program on Semiconductor Synthetic Biology (SemiSynBio, SSB) which provides seed funding for research on synthetic biology as it relates to computing/information processing [23]. While this initiative seems ripe for a possible SRC-NSF follow-up joint program, it does require DA tools that can aid the development of SSB systems for any progress to be achievable.

## 4.3 Design Automation of *Things* (DAoT)

The new markets likely to benefit from the DA body of knowledge share several distinguishing features:

- High potential for disruptive impact;
- High structural complexity and deep abstraction hierarchies;
- Heterogeneous components or *things*, representing physical, cybernetic or information environments.

To leverage the basic tenents of Design Automation, as well as the skills of DA professionals, in these new markets, we envision a broader field focusing on the *Design Automation of Things* (DAoT). In this context, immediate attention areas for "things" are as follows:

- *Medical Technology:* traversing the areas of wearable medical and personal-health devices, *in vivo* sensed or sampled tissues for personalized diagnosis, prosthetics, synthetic biology from organs to individual cells, etc.;

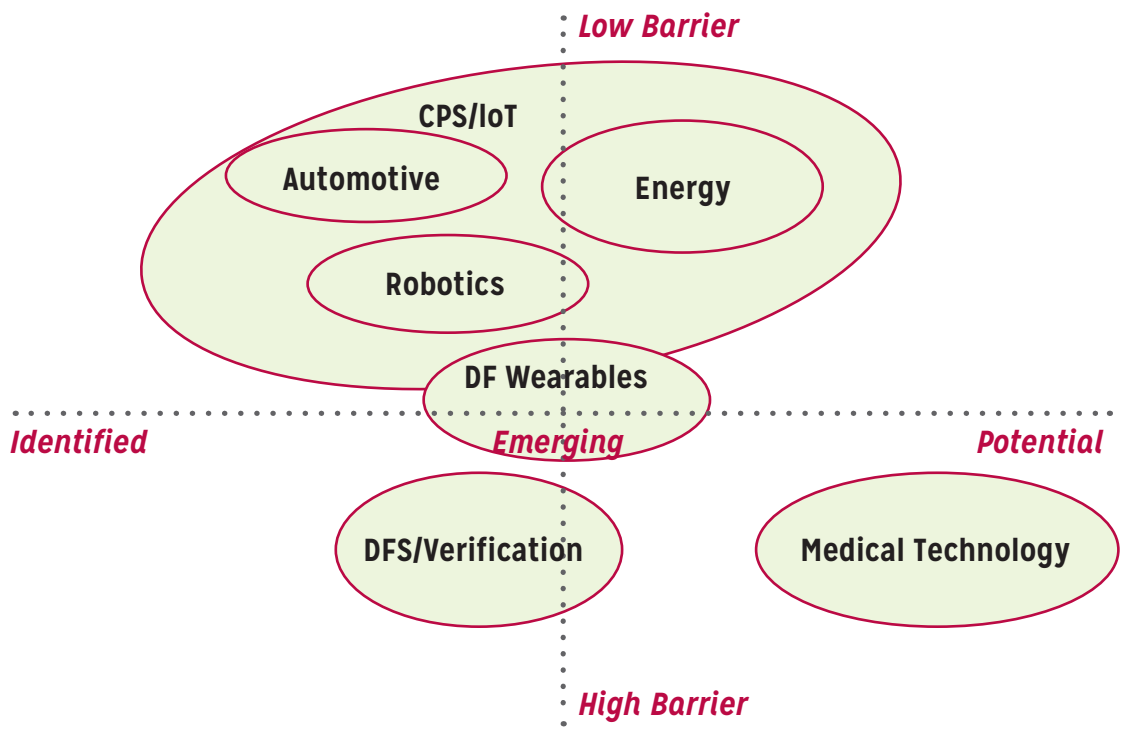

Figure 7: New markets where the DA body of knowledge and skills can be tranferred.

◗ **Cyber-physical Systems (CPS):** including cloud-connected (smart) computing devices responsible for data acquisition and processing, such as smartgrid components and home entertainment systems - same as "things" in the IoT context [24] that include sub-fields such as;

- *Automotive and Aerospace Systems:* utilizing electronics for sensors and rangers (radars and lidars), device and subsystem controllers, local networks, long-range communication, and navigation;

- *Industrial and Robotics Systems:* such as electronics used in manufacturing-control, facilities maintenance, and mass transit systems;

- *Energy Systems:* from electronics used in the oil and gas industry (e.g., equipment used in exploration and refining), smart energy distribution networks (i.e. Smart Grids), and energy transportation,

◗ **Design for Security (DFS):** cyber-secure systems such as those used in biometrics, secure communications and control, home monitoring, remote medical diagnosis, telesurgery, etc.;

If properly developed, this field is poised to enjoy a *technical* edge and promises deep synergies with a variety of adjacent fields.

**From a technical perspective,** DAoT can address complex problems with a focus on *engineering design* in a *hierarchical* and *unified manner*, generalizing single-instance construction and automating demanding work.

**From a community perspective**, the approach taken by EDA has provided a concrete, successful framework of engaging academic research, industry, and government. Traditional academic EDA research implemented theoretical developments in practical systems through collaboration with industrial, research lab, or federal support (e.g., via programs sponsored by SRC and NSF). A prosperous DAoT field, focusing on "things" of *broad practical relevance* that satisfy *clear need* and have revolutionary impact, can leverage the successful EDA model and maintain progress at rates comparable to those of Moore's law.

**The DAoT approach to new markets** can help alleviate market pressures from the peaks and valleys of the hype curve illustrated in Figure 5. In return, the (somewhat mature) semiconductor industry will be able to *amortize* R&D among many such markets and exploit economies of scale.

## 5  Cross-Cutting Challenges

As Electronic Design Automation enters the post-scaling era, several challenges permeate multiple thrusts. The lack of appropriate design and EDA abstractions for emerging technologies and new markets is one, mirroring the early days of IC design. Moreover, the potential of new computer architectures has long been symbiotic with the capabilities of EDA. Further, EDA is showing signs of stagnation similar to those in the power-electronics industry with a relatively static and aging workforce. It took a dramatic power-loss event and considerable investment to revitalize the field, ushering the smart-grid era. A comparable disruption in EDA could lead to failed chip designs, loss of competitiveness in major industries, and stunted growth. Thus, the next sections discuss these cross-cutting challenges and related recommendations.

### 5.1  Effective Abstractions, Design Metrics, and EDA Benchmarks

A research focus on the development of EDA abstractions, design metrics, and representative benchmarks, as has been highlighted in each of the previous three sections, may seem unexpected, but is of critical importance. In areas where effective abstractions have been developed, dramatic progress has been possible. In situations where research has stagnated, better abstractions, or better refined metrics and benchmarks has often been a catalyst for improved progress. In areas where abstractions, metrics, and benchmarks have been absent, progress has been slow. This can be illustrated by analog circuit design–an area of many opportunities, hampered by a shortage of effective automated tools (Section 2.2). Looking forward to emerging challenges for Design Automation, we foresee important challenges and broad opportunities to develop new EDA abstractions, design metrics, and benchmarks. The abstractions for traditional silicon design flows have started to develop cracks as Dennard scaling broke down, dramatically increasing the amount of information that must be exchanged across abstraction layers. Furthermore, single-abstraction optimizations are known to leave much room for improvement, whereas multiobjective optimizations that can span several nearby levels of abstraction have shown consistent improvements. In newer areas, known abstractions are often rudimentary (if they exist at all) or adopted from existing abstractions from other domains, sometimes aptly sometimes not. Below, we illustrate sample needs for abstractions where current approaches are inadequate or missing.

- System-level abstractions appropriate for extreme-scale IC designs that can be effectively validated (Section 2.2).

- Abstractions for emerging technologies that enable full-system design flows (Section 3.1).

- Abstractions for new technology markets including biology, cyber-physical systems, and hardware security (Section 4.1).

Investment into appropriate abstractions for these domains can catalyze the development of prototype EDA flows and enable new types of systems. Benchmarks and computational experiments can then help evaluate and compare such systems. With abstractions in place, design metrics can capture application needs and drive research in optimization. A universal agreement on the best metrics for a particular problem is not necessary, but it is important to establish common ground. A research infrastructure based on EDA abstractions, design metrics, and benchmarks will support effective reuse of results and best practices, as well as crosspollination of ideas between different research avenues. An iterative process, where EDA abstractions, design metrics, and benchmarks are continually re-evaluated and refined can supercharge research efforts and make them more relevant to the industry. This is happening in traditional EDA research and is the sign of an active, vibrant research community. Research in emerging areas should follow this path.

## 5.2 EDA Synergies With Computer Architecture

For many years, general-purpose microprocessors, ranging from the smallest embedded processors to workstation-, server-, and supercomputer-class processors, have been the main driving application within the EDA industry. In the early part of the millenium, multimedia processors and general purpose graphics processing units (GPGPUs) gained significance, and in the last five years have culminated with even embedded palmtop computing applications seeing systems-on-chip (SoC) architectures (e.g., including several embedded CPUs with a tightly integrated GPU) becoming more prominent. Such high-volume products are typically at the leading edge of fabrication technology and design complexity, putting the focus of EDA support to target these IC designs. As such, Computer Architecture and Design Automation are affected by many of the same market forces, technology roadmaps, and application trends. This context motivates us to consider how research in EDA and Computer Architecture interact.

First and foremost, the end of Dennard scaling has limited clock rates achievable with reasonable power and cooling constraints. This has, in turn, constrained serial performance, and put an emphasis on "energy first" philosophy [25]. However, the desire to continue Moore's Law has led to research in new methods to increase computational performance. The most prevalent method, driven by the architecture community, has been the increase of core-level parallelism. Architectures have quickly graduated beyond single-core superscalar cores to hyperthreading and eventually homogeneous multi-core products. A second trend is for emerging massively parallel chip-multiprocessor systems (i.e., many-core processors) to become increasingly heterogeneous. Such systems include embedded/small cores, traditional superscalar cores, GPU cores with hundreds of threads, and dedicated special purpose cores. These systems can be configured or tuned for the target application domain.

Just as EDA is finding itself in an environment similar to the days prior to Dennard scaling, Computer Architects are rethinking the fundamental concepts of architecture for similar reasons. Both fields must adapt to a new frontier where performance and power advantages must look inward for innovation and must contend with considerable challenges of leaky and unreliable circuit elements [26]. Additionally, these fields must address the increased complexity of hybrid fabrication technologies currently driven by market trends pushing architectures to integrate more, highly-dense, non-volatile storage technologies on-chip [27]. As a result, advancements in post-scaling EDA and Computer Architecture will depend on improving the quality and capability of EDA tools alongside the fundamental design of computing architectures. These efforts will be most effective if EDA and architecture challenges are addressed with combined knowledge of the two fields.

**Key challenges in Computer Architecture** [25] and their implications to Design Automation are outlined below.

- *Smart Sensing and Computing* devices will perform data processing at the "sensor" level–with extremely low power constraints. Design Automation must consider the integration of multiple device technologies (to support different types of sensors), with the design challenge being holistic in nature, rather than focusing on one part of a system.

- *Portable Edge Devices* will expand on an already thriving area of embedded systems. From smart phones to tablets, and extending to more forward looking devices (e.g., Google Glasses), there will be a need for tight integration between multiple conventional synthesized processors (such as the ARM cores in Apple's iPhone processor line), graphics acceleration pipelines, and a multitude of software radios. These devices can be viewed as heterogenous computing platforms, with a need to continually expand and update features, while maintaining a large degree of software portability.

- *Cloud Servers* are expected to provide a large amount of "bulk computation" at modest costs. These systems will likely entail large numbers of parallel processors, which can be reconfigured and reassigned as computing demands change, while maintaining low power. These processors will contrast sharply with conventional *single-user high-performance* chips.

ITRS describes chip power as a key design concern for the near future. Specialization may be a key aspect of effective power optimization; in many application areas, customized circuits outperform general-purpose processors by orders of magnitude. An abundance of silicon area, coupled with limited power, may result in large numbers of specialized cores and co-processors on a single chip. By turning these cores on and off as needed, the next generation of chips may be able to be both faster and less power-hungry. Innovation in Design Automation tools will help handling increased size and complexity. An open question, however, is whether specialized designs can achieve sufficient volume to make them cost effective. Reconfigurable hardware (such as FPGAs coupled with conventional low-power processor cores) may provide the performance levels required–but through reconfigurability, they can address a wider range of applications, leading to lower unit costs.

**Changes to the Programming Model:** Like changes to traditional EDA, the end of traditional scaling has led the Computer Architecture community towards massive parallelism as a way to achieve increased performance without exploding power-density of sub 90nm technologies. Some envision 1000-core chip designs [28], but their performance will be limited by Amdahl's Law [29]. For applications with abundant parallelism (e.g., graphics processing), there are highly successful, massively parallel solutions. While the scientific community has leveraged parallel systems for decades, the tuning of applications for these systems is often tightly coupled with the hardware to achieve the best possible performance gains. Other examples include parallel processing for data centers and cloud systems for multi-program workloads and independent virtual machines [30]. Outside of these application areas, the prospects for massive parallelism are more limited and can be satisfied by a limited number of parallel cores. However, in all of these cases significant bottlenecks continue to exist. The most common is efficient access to data through limited bandwidth memory systems. The next is the limited capability of software designers to most efficiently utilize resources. These challenges have similarities with EDA problems at the extreme-scale, particularly in the development of efficient system-level design flows.

Computational complexity theory [31] shows that the algorithm selected to accomplish a task often drives the performance of the solution more than the hardware on which the task runs. For a static architecture, this limits the reasonable algorithmic choices for most applications-and a large percentage of computationally efficient algorithms have limited parallelism. A "1000-core" chip, in which the individual cores are customized for a wide range of different tasks, resonates with many researchers, which motivates domain-specific hardware optimizations [32]. Hence, system-level collaboration between EDA researchers and computer architects can make processors more efficient, as illustrated by early industry efforts such as Tensilica (acquired by Cadence). Today, when homogeneous parallelism is no longer the sole target and design complexity must increase, such collaboration is particularly attractive.

## 5.3 Maintaining a Workforce of Future Design Automation Experts

At the start of the semiconductor revolution, there was significant government funding, both for the underlying technology, and also for the software tools to enable design. There was an effective eco-system: the funding enabled a great deal of fundamental research at a time when there was no established "industry" presence to shoulder the burden. Academic research served to train the first generations of industrial pioneers and entrepreneurs. The successful launch of new companies and the creation of new jobs could be considered as a return on the initial funding investment.

The landscape today has changed considerably. There are now many successful semiconductor companies involved in both manufacturing and design tool development. Given the industry presence, government funding agencies have reduced their support, placing more emphasis on areas that are considered high-risk/high-reward. These changes have placed a number of stresses on the research eco-system. Funding of academic research from industry groups is highly competitive, and can be narrowly focused, while government funding is scarce. It has become difficult for academics to fund graduate research, resulting in fewer professionals entering into the industry, and the potential of a severe shortage of talented workers in the future. Further, many of the best and brightest students are being lost to other research areas. While a significant hike in government funding would be quite welcome, moderate increases are more realistic. The workshop participants offered a number of suggestions that can help adapt to these challenges [33].

- ◗ Pooling of educational resources, and amortizing the effort of developing educational material, is a worth-while goal. Initial experiments with a design-automation centric MOOC course appeared successful, introducing the research field to students who might not have otherwise been aware of it.

- ◗ Activities such as the SIGDA Design Automation Summer School were also well received. Relatively few universities have the depth of faculty (or scheduling ability) to offer a full range of courses in the topics covered in design automation. Short, intensive courses may have broad impact.

- ◗ Material developed for courses such as those presented through the summer school could be developed into more formal material, to help jump-start current design automation researchers into new fields.

## 6 Conclusions and Recommendations

The development of future generations of increasingly capable electronic computing systems has become a driving force of society. However, traditional technology scaling, and along with it, the technology driven advances in performance and reductions in power have already reached an end. We refer to this post-traditional scaling for electronics as the *extreme-scale* era. In extreme-scale design automation, the tools must re-emerge as the driver to achieving advancements in system design criteria including performance, power-density, reliability of design, integration of new electronics technologies, and even towards integrating market driven support for security and mobility. *Now, more than at any time in its history, the success of the EDA field is crucial to enabling further developments in next generation systems.* In the following sections we discuss recommendations and actionable items from the workshop discussions on each of the three focus areas.

## 6.1 Challenges and Opportunities in Extreme-scale EDA

In order to continue the push to enabling systems with technologies under 10nm considerable advancement is still required in EDA. These directions are being pursued primarily by industry with limited investment by governmental agencies in academic research. These important areas of emphasis in traditional EDA include:

- Traditional aspects of advanced process technologies, such as process complexity.

- Building reliable systems with unreliable components.

- Interactions between design-rule complexity and layout complexity.

- Scaling in tool capacity + scaling ICs to extreme scale and mega-structures.

- Custom analog, semi-automated analog, and mixed-signal.

Amongst the most important advancements are the scaling of tool capacity, particularly in validating and verifying systems of extreme-scale integration.

Key drivers of progress sometimes conflict with each other. To solve problems related to process uncertainty, more information is required to be communicated across the layers of the EDA flow. However, this additional information exchange works against the scale of design. The desire to handle the complexity of design continues to raise the level of abstraction for design, which has the potential to further decrease the quality of the final design. These trends call for investment in a targeted program towards the *effective new abstractions* for design that reduce information communicated across the layers while *raising the level of abstraction* of design and encouraging *effective validation and verification* of the final system. Such directions must not rehash approaches from other domains (such as software) but, rather re-invent abstractions for effective system-level design and verification at the extreme-scale.

A second direction is to invest in closing the gap between EDA tool capability and technology for ›80nm technologies. It has become clear that post-80nm technologies are only targeted by a small subset of the integrated circuit designs that require the level of integration possible in such designs. These designs are dominated by commodity computer processors and other general purpose chips (such as FPGAs) that can effectively (and have reason to) amortize the upfront costs of manufacturing at these leading-edge technologies. In comparison, many design starts now target well-established technologies including those at the 130nm node, marking the tail end of what is possible with traditional scaling. As EDA continues to pursue sub-10nm design, extracting the most out of 130nm technologies has not been a focus. Leveraging inexpensive use of established technology nodes has been identified as a promising direction recently by many SoC IP companies [34]. As Moore's law no longer gets driven by technology scaling, advancement is possible in older, less aggressive technologies through innovation in application or system-level design. A targeted investment in *revolutionary EDA techniques* that maximize what can be obtained in pre-80nm technologies is needed. Such approaches can provide superlow power and cost solutions for existing and emerging markets. While critical, this research must overcome the notion of EDA for pre-80nm as a solved problem. A strategic approach of a possible NSF/DoD program is recommended for this challenge.

Finally, lack of active research in traditional areas may undermine competence in graduate-level EDA education, and therefore in abilities to attack emerging and related problems. This is particularly relevant to the domestic need and NSF's mission to ensure a high-tech workforce, in addition to NSF's mission to support research. The industry should also be a major stakeholder.

## 6.2 Supporting Design of Hybrid and Post-CMOS Technologies

The future advancements of electronics will continue to rely on new developments in technologies. While there is still much value that can be extracted from silicon-CMOS, other semiconductor and nano-scale technologies still maintain considerable untapped potential. Some traditional EDA research can be adapted to emerging technologies, but alone such approaches will not be sufficient without further investment. A further challenge is how EDA research can progress for technologies associated with strategic uncertainty.

*We propose research in EDA strategies specifically targeted to diverse and multiple-use techniques including abstractions, optimization methodologies, and techniques to address process-specific design considerations.* Eventually, as certain emerging technologies start to show more demonstrable promise, these abstractions and methodologies can be refined and customized to reflect the specific features of those technologies. Furthermore, advances in the use of emerging technologies and identifying promising candidates for CMOS replacement cannot be achieved without proper device modeling and simulation technologies. *Therefore, an indispensible component for achieving identifiable outcomes in the post-CMOS era is the development of device models that can be used for characterizing emerging architectures and accompanying simulation tools that can enable design exploration and optimization.*

## 6.3 Design Automation of Things

Design automation principles have the potential for transformative benefits in other application domains. Given the relative maturity of design automation for electronics, now is a natural time to apply its methodologies into other application areas. There is an immediate need for increased investment in design automation for cyberphysical systems and cyber-secure systems. A longer term development strategy for investment is required for farther fields. Biological and medical technologies have been identified as appropriate initial focus areas. The exploration strategies for biology and medical technologies can serve as a model to address further focus areas as they are identified.

**Design Automation for Cyber-physical Systems and Cybersecurity**

The need for applying design automation techniques to both cyber-physical systems and cyber-secure systems have been identified as immediate needs for direct support. Existing visioning for these targeted research efforts identified EDA as an important research need. Unfortunately, NSF programs in these areas do not enumerate this element in their solicitations.

Cyber-physical systems, or the Internet of Things, is a logical first step in applying design automation outside of the traditional electronics scope. Design automation for CPS/IoT has been identified as early as 2008 by the CPS community [22] and reinforced by the EDA community (including these workshops) as a critical component for transformative progress.

Cyber-secure systems and, in particular, hardware security indicate a second immediate need for directed investment in the design automation of things. The recent SaTC call in hardware security, STARSS [21], focuses on design of cyber-secure systems but largely ignores design automation techniques required to make these design techniques scalable.

The importance of addressing the need for focused EDA research in CPS and cyber-secure systems is further supported by discussion with NSF leadership and the CCC workshop community. Thus, we recommend to establish direct links between EDA research and the existing programs in CPS and STARSS to enumerate direct funding efforts dedicated towards design automation research activities supporting these areas. This can be achieved in the short term through targeted "Dear Colleague Letter" calls and in the long term by establishing possible stand-alone programs focused on these topics.

**Design Automation for Far Fields**

To develop transformative application of EDA principles to different problem domains including those entirely removed from the traditional electronics market, a new *learning community* of researchers is needed. While several EDA researchers have made the transition to new markets successfully, a prosperous community requires more than individual efforts. We need the *creation of patient capital for exploration* in an area that may take a relatively long time to establish itself and provide tangible results. Without such a framework, it will be difficult to keep members interested and prevent attrition. Past efforts that have featured new markets as possible applications of DA in established EDA venues (either by special sessions or tutorials) have produced mixed results. We need a *distinct* research community that unifies all DAoT efforts under a single umbrella.

|  | **Electronics:** Hybrid CMOS with Emerging Technologies | **New Markets:** Cyber-physical, Cyber-secure, and Bio-medical Technologies |
|---|---|---|
| **Traditional EDA Tool-kit** | Immediate Need: EDA for scaled CMOS + product ready tech | Immediate Need: EDA applied to *near fields* – automotive, robotics, and energy |
|  |  |  |
| **EDA Approaches on Big Data** | Transformative: *Big data* research–system level design and verification | Transformative: EDA big-data methodologies applied to *far fields* – synthetic bio, systems bio, medical devices |

*Figure 8: An overview of the challenges for the EDA field to enable the technologies of 2025 and beyond. EDA will continue to remain relevant in enabling the newest technologies and has already started to be leveraged in related problem domains with similar challenges to the design of electronics. While these directions remain relevant, the research that will transform EDA for 2025 is the investment in (1) big-data approaches for electronics challenges of system-level design, emulation and verification, and (2) application of EDA approaches of abstraction, optimization, validation to non-traditional electronics (electronics+) fields requiring the development of appropriate abstractions, operating on large data-sets and non-linear optimization needs.*

While initially a workshop would be required to understand where the problems are and how they could be addressed, in the long term a sustained event (or network thereof) is what will drive the community further.

Specific recommendations include:

**1.** A network of workshops focused on the process of using the proposed DAoT tools, rather than the results achieved by them. The same topic can be described by a potential customer, a field expert, and a DA expert, thereby uncovering the "why" and "how," in addition to "what" is done.

**2.** An annual Gordon-like/Dagstuhl-like conference [35, 36] where all participants present and actively define the field by brain-storming in breakout discussion, generating outcome documents, and continually guiding the future of the field. The community of practice will need to group junior and senior researchers in a framework similar to a Laureate Forum [37], thereby helping emerging scientists make progress in a new field while burnishing their own individual profile as independent researchers.

Such a new community requires the sustained support of relevant professional societies (Association for Computing Machinery, Institute of Electrical and Electronics Engineers), but most importantly needs the type of patient capital for exploration that has traditionally come from federal and industrial funding agencies (NSF, DoD/DARPA, or SRC).

## 6.4 Recommendations and Next Steps

An investment in EDA proposed before should be strategically positioned to have the greatest positive impact. Electronic Design Automation has matured considerably in 40 years, but challenges remain in extreme-scale design for sub-10nm technology nodes with more than $10^{15}$ elements. Such electronic designs are becoming increasingly hybrid in nature. Scalable methodologies for inclusion of new post-CMOS non-silicon technologies into design require effort. While a "Mead and Conway" approach can provide insights into abstractions, new and effective abstractions for non-silicon technologies may bring significant benefits. Compelling roadmaps must be developed based on research progress benchmarked against technology agnostic-metrics. While the expectations should not be to immediately exceed CMOS to be successful, technologies that have fatal and insurmountable limitations must give way to those that have the potential for success. Finally, EDA must branch out from electronics and specifically from IC design, alone. Design automation is a *big-data* discipline that can provide considerable advancements in other important fields such as manufacturing and medical science and technology.

The intersection between EDA and future technologies reaching beyond electronics should be explored. This intersection is illustrated in Figure 8.

**We offer the following recommendations:**

### TRADITIONAL EDA
Research in traditional design automation is still important and should be supported at least at current levels, if not with a nominal increase. This research will help close the productivity gap between what is possible with scaled CMOS and what can be realized with tools at post-scaling technologies. Additionally, the ability to develop new EDA tools to match a slightly altered design context is routinely required in modern VLSI Design research to evaluate new circuit techniques.

### EXTREME-SCALE EDA
Targeted investment in EDA to support < 10nm technologies is required beyond existing funding levels. A new program in the *big-data* research space on extreme-scale EDA is required to address challenges of system-level design, verification, and reliability to continue further scaling to 2025. Joint SRC/NSF programs will foster academic and industrial collaboration to address this ongoing challenge.

### HYBRID POST-CMOS ELECTRONICS
Revisiting a program to promote abstractions, benchmarks, and metrics for emerging technologies is necessary to enable the incorporation of non-CMOS technology into designs. The original NSF Nanoelectronics for 2020 and Beyond (NEB) program had this direction included as a goal. Perhaps a similar program must be developed specifically for non-CMOS technologies. As part of the deliverables, EDA abstractions, design metrics, benchmarks, and results of computational experiments should be treated as real and significant research results.

### DESIGN AUTOMATION OF THINGS
Design automation techniques must be considered for both near and far fields from electronics. These techniques have the potential to revolutionize other industries based on the developments made from 40 years of investment in electronics. Supported by NSF leadership, we recommend a "Dear Colleague Letter" be prepared to support directed research for design automation for CPS and STARSS. For farther fields such as biological and medical technologies, exploratory research should be considered. For example, an Emerging Frontiers in Research and Innovation (EFRI) topic on design automation for synthetic biology in collaboration with the SRC SSB [23] program could be a good first step. However, a long term thrust in DAoT that includes CPS and cyber security will also address important innovation at the intersection of IoTs, personal(ized) healthcare, and wearable systems..

We believe that, by following up on these recommendations, the Electronic Design Automation community (and Design Automation more broadly) will extend leadership in established as well as new markets, while being an exciting and essential part of future technological development.

## References

[1] R. H. Dennard, F. H. Gaensslen, V. L. Rideout, E. Bassous, and A. R. LeBlanc, "Design of ion-implanted MOSFET's with very small physical dimensions," in *IEEE Journal of Solid-State Circuits*, vol. 9, no. 5, pp. 256-268, Oct. 1974.

[2] G. E. Moore, "Cramming More Components onto Integrated Circuits," *Electronics Magazine*, pp. 114-117, Apr. 1965.

[3] C. A. Mead and L. A. Conway, *Introduction to VLSI Systems*, Addison Wesley, 1980.

[4] R. A. Rutenbar, "DAC at 50: The Second 25 Years," *IEEEE Design & Test of Computers*, vol.31, no.2, April 2014, pp. 32-39.

[5] S. Rambo, "Cadence Expands Its Systems Definition," *EE Times*, 3/13/2014.

[6] International Technology Roadmap for Semiconductors, [Online]. http://www.itrs.net/reports.html.

[7] E. Lehtonen and M. Laiho. "Stateful implication logic with memristors," in *Proceedings of the 2009 IEEE/ACM International Symposium on Nanoscale Architectures* (NANOARCH '09), Jul. 2009, pp. 33-36.

[8] J. Watrous, "On one-dimensional quantum cellular automata," in *Proceedings of 36th Annual Symposium on Foundations of Computer Science*, pp. 528-537, Oct. 1995.

[9] R. E. Bryant, "Symbolic Boolean Manipulation with Ordered Binary Decision Diagrams," *ACM Computing Surveys*, vol. 24, no. 3, pp. 293-318, Sept. 1992.

[10] A. de Geus, "The Greatest" Tech-Onomic Push-Pull in Human History," *IEEE Design & Test of Computers*, vol.31, no.2, April 2014, pp. 9-12.

[11] R. K. Brayton and J. Cong, "NSF Workshop on EDA: Past, Present, and Future (Part 1)," *IEEE Design & Test of Computers*, vol. 27, no. 2, pp. 68-74, 2010.

[12] R. K. Brayton and J. Cong, "NSF Workshop on EDA: Past, Present, and Future (Part 2)," *IEEE Design & Test of Computers*, vol. 27, no. 3, pp. 62-74, 2010.

[13] K. Flautner, "The State of the Future," Keynote Talk, CEDA Luncheon at DAC, 2014.

[14] L. Stok, "The Next 25 Years in EDA: A Cloudy Future?," *IEEE Design & Test of Computers*, vol. 31, no. 2, April 2014, pp.40-46.

[15] M. M. Ozdal, Ch. Amin, A. Ayupov, S. M. Burns, G. R. Wilke, and C. Zhuo, "An Improved Benchmark Suite for the ISPD-2013 Discrete Cell Sizing Contest," in *Proceedings of International Symposium on Physical Design*, Mar. 2013, pp. 168-170.

[16] M. C. Kim, N. Viswanathan, Z. Li, and C. J. Alpert, "ICCAD-2013 CAD Contest in Placement Finishing and Benchmark Suite," in *Proceedings of International Conference on Computer Aided Design*, Nov. 2013 pp. 268-270.

[17] V. Yutsis, I. Bustany, D. G. Chinnery, J. R. Shinnerl, and W.-H. Liu, "ISPD 2014 Benchmarks with Sub-45nm Technology Rules for Detailed-routing-driven Placement," in *Proceedings of International Symposium on Physical Design*, Mar/Apr 2014, pp. 161-168.

[18] D. Drako, "The Big Data Advantage," Opening Talk for the Panel on Conquering Multi-site Design, Panel Participants: D. Nikolic, S. Sikand, C. Leung, A. Lewis, in *Proceedings of the Design Automation Conference*, June 2, 2014.

[19] "Gartner's 2013 Hype Cycle for Emerging Technologies Maps Out Evolving Relationship Between Humans and Machines," Press Release, August 2013 [Online]. [Online] Available: http://www.gartner.com/newsroom/id/2575515. [Accessed: 20-Jun-2014.]

[20] "Research Needs for Secure, Trustworthy, and Reliable Semiconductors." [Online] Available: https://www.src.org/calendar/e004965/sa-ts-workshop-report-final.pdf. [Accessed: 2-Jun-2104].

[21] "Secure and Trustworthy Cyberspace: Secure, Trustworthy, Assured and Resilient Semiconductors and Systems (SaTC: STARSS)." [Online]. [Online] Available: http://www.nsf.gov/pubs/2014/nsf14528/nsf14528.htm. [Accessed: 2-Jun-2014].

[22] "Cyber-Physical Systems: Executive Summary." [Online]. Available: http://varma.ece.cmu.edu/summit/CPS-Executive-Summary.pdf. [Accessed: 2-Jun-2014].

[23] "Synthetic Biology Ramps Semiconductors." [Online]. Available: https://www.src.org/newsroom/press-release/2013/521/. [Accessed: 2-Jun-2014].

[24] G. Lowe, "Driving the Internet of Things," *IEEE Design & Test of Computers*, vol. 31, no. 2, April 2014, pp. 22-27.

[25] M. Hill, et. al., "21st Century Computer Architecture: A Community White Paper," May 2012.

[26] S. Borkar, "Designing Reliable Systems from Unreliable Components: The Challenges of Transistor Variability and Degradation," *IEEE Micro*, vol. 25, no. 6, pp. 10-16, Nov. 2005.

[27] S. Tehrani, "Recent Development and Progress in Nonvolatile Memory for Embedded Market," Keynote Address to the asQED Symposium (Slides), 2012. [Online] Available: http://www.asqed.com/English/Archives/2012/Misc/Tehrani Keynote ASQED web.pdf. [Accessed: 10-Jun-2014].

[28] J. Torellas and M. Oskin, "Workshop on Advancing Computer Architecture Research (ACAR-1)/Failure is Not An Option: Popular Parallel Programming," August 2010. [Online] Available: http://www.nitrd.gov/nitrdgroups/images/5/58/ACAR1 REPORT.pdf. [Accessed: 15-Jun-2014].

[29] M. Oskin and J. Torellas, "Workshop on Advancing Computer Architecture Research (ACAR-2)/Laying a New Foundation for IT," September 2010. [Online] Available: http://www.nitrd.gov/nitrdgroups/images/1/14/Workshops On Advancing Computer Architecture Research.pdf. [Accessed: 15-Jun-2014].

[30] L. A. Barroso, J. Clidaras, and U. Hölzle, "The Datacenter as a Computer: An Introduction to the Design of Warehouse-Scale Machines," *Synthesis Lectures on Computer Architecture (2nd ed.)*, Morgan & Claypool Publishers, 2013.

[31] M. Sipser, "Introduction to the Theory of Computation," 3rd ed., International Thompson Publishing, 2012.

[32] J. Cong, G. Reinman, A. T. Bui, and V. Sarkar, "Customizable Domain-Specific Computing," *IEEE Design & Test of Computers*, vol. 28, no. 2, pp. 6-15, 2011.

[33] I. Bahar, A. K. Jones, S. Katkoori, P. H. Madden, D. Marculescu, and I. L. Markov, "Scaling" the impact of EDA education - Preliminary findings from the CCC workshop series on extreme scale design automation," in *Proceedings of Microelectronic Systems Education (MSE)*, June 2013, pp.64-67.

[34] H. Yassaie, "The Great SoC Challenge (IP To The Rescue!), Imagination Technologies Ltd." [Online] Available: https://dac.com/events/2014-06-02. [Accessed: 2-Jun-2014].

[35] "Gordon Research Conference." [Online] Available: http://www.grc.org/. [Accessed: 27-Apr-2014].

[36] "Schloss Dagstuhl: About Dagstuhl." [Online]. Available: http://www.dagstuhl.de/en/about-dagstuhl/. [Accessed: 27-Apr-2014].

[37] "Abel, Fields and Turing Laureates Meet the Next Generation–Heidelberg Laureate Forum." [Online]. Available: http://www.heidelberg-laureate-forum.org/. [Accessed: 28-Apr-2014].

**Notes:**

______________________________________________________________________

______________________________________________________________________

______________________________________________________________________

______________________________________________________________________

______________________________________________________________________

______________________________________________________________________

______________________________________________________________________

______________________________________________________________________

______________________________________________________________________

______________________________________________________________________

______________________________________________________________________

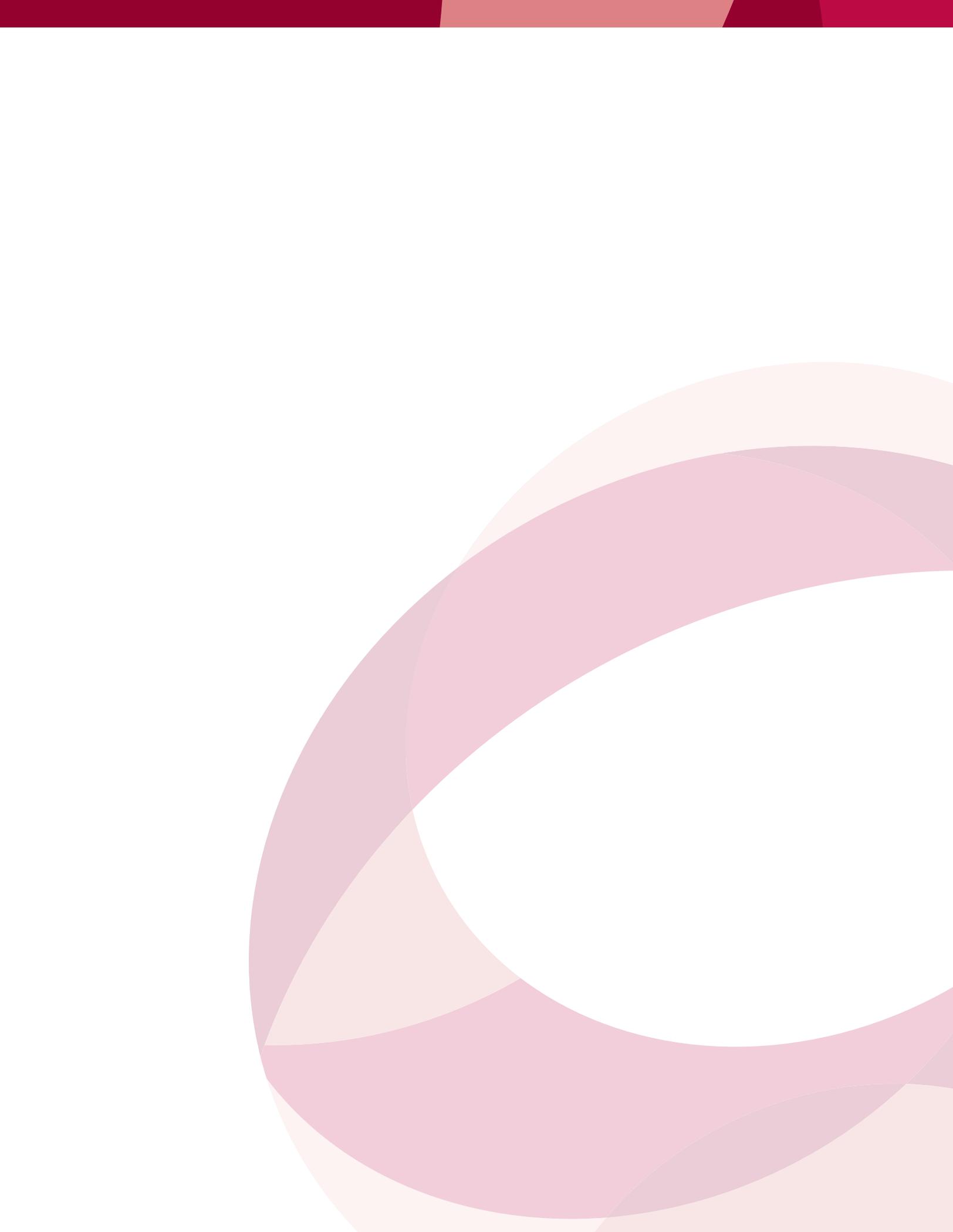

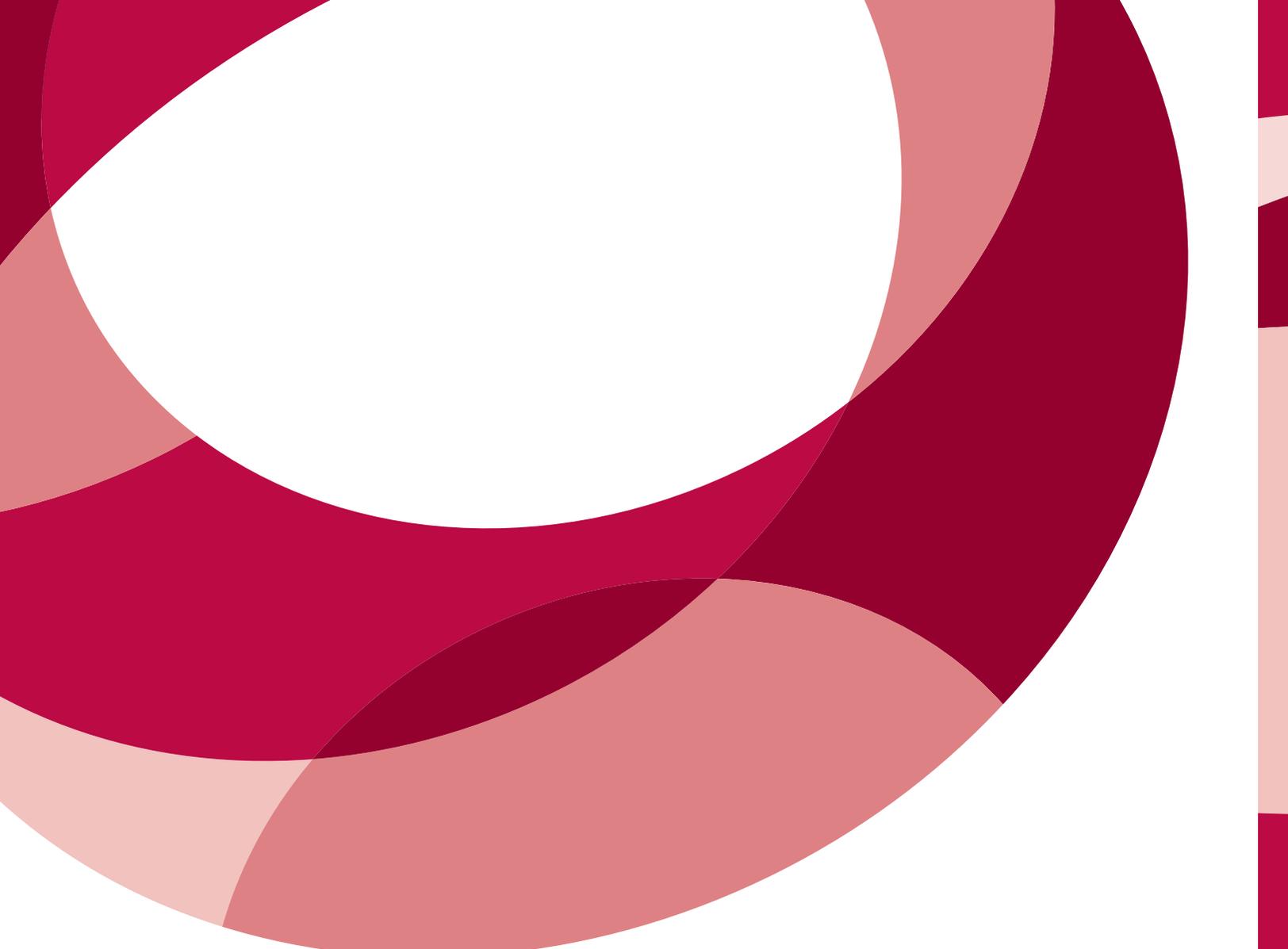

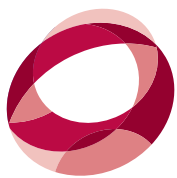
**CCC**
Computing Community Consortium
Catalyst

1828 L Street, NW, Suite 800
Washington, DC 20036
P: 202 234 2111  F: 202 667 1066
www.cra.org  cccinfo@cra.org